\documentclass[a4paper,11pt]{article}
\usepackage{pos}

\title{Astrophysical searches of ultralight particles}

\author*[a,b]{Tanmay Kumar Poddar}

\affiliation[a]{Theoretical Physics Division, Physical Research Laboratory,\\
  Shree Pannalal Patel Marg, Ahmedabad-380009, India}

\affiliation[b]{Department of Theoretical Physics, Tata Institute of Fundamental Research,\\
Homi Bhabha Road, Mumbai-400005, India}

\emailAdd{tanmay@prl.res.in}
\emailAdd{tanmay.poddar@tifr.res.in}

\abstract{The Standard Model of particle physics is a $SU(3)_c\times SU(2)_L\times U(1)_Y$ gauge theory that can explain the strong, weak, and electromagnetic interactions between the particles. The gravitational interaction is described by Einstein's General Relativity theory which is a classical theory of gravity. These theories can explain all the four fundamental forces of nature with great level of accuracy. However, there are several theoretical and experimental motivations of studying physics beyond the Standard Model of particle physics and Einstein's General Relativity theory. Probing these new physics scenarios with ultralight particles has its own importance as they can be a promising candidates for dark matter that can evade the constraints from dark matter direct detection experiments and solve the small scale structure problems of the universe. In this paper, we have considered axions and gauge bosons as light particles and their possible searches through astrophysical observations. In particular, we obtain constraints on ultralight axions from orbital period loss of compact binary systems, gravitational light bending, and Shapiro time delay. We also derive constraints on ultralight gauge bosons from indirect evidence of gravitational waves, and perihelion precession of planets. Such type of observations can also constrain several particle physics models and are discussed.}

\FullConference{%
  41st International Conference on High Energy physics - ICHEP2022\\
  6-13 July, 2022\\
  Bologna, Italy
}

\begin{document}
\maketitle

\section{Introduction}
The Standard Model (SM) of particle physics is a very successful theory as it can explain the strong, weak, and electromagnetic interactions in nature. The fourth fundamental interaction is the gravitational interaction which can be explained by Einstein's General Relativity (GR) theory. Both of these theories can explain the known fundamental interactions between the particles very well with impressive level of accuracy. However, there are theoretical and experimental motivations of studying physics Beyond the Standard Model (BSM) and Einstein's theory of gravity. For example, in SM of particle physics, neutrinos are massless. However, several solar, atmospheric, and reactor neutrino experiments have confirmed that neutrinos are massive. The galactic rotation curve, Cosmic Microwave Background (CMB) spectrum, and structure formation have confirmed the existence of non luminous (Dark) Matter (DM) in the universe, but there is no such DM candidate in SM. There are other issues which cannot be explained by the known SM of particle physics such as matter-antimatter asymmetry of the universe, strong CP problem, hierarchy problem, gravitational interaction etc. There are few shortcomings in Einstein's GR theory as well such as singularity problem, cosmological constant problem etc. In this paper, we have considered ultralight pseudoscalar axions, and $U(1)$ vector gauge bosons that can solve some of the new physics scenarios and their possble searches through astrophysical experiments. These ultralight particles can be promising candidates for Fuzzy Dark Matter (FDM) and fifth force. Unlike Weakly Interacting Massive Particle (WIMP) model, the FDM model can explain the small scale structure problem and evade the constraints from dark matter direct detection experiments.

Axion was first proposed by Peccei and Quinn (1977) as the solution of the strong CP problem. It is a pseudo Nambu Goldstone Boson (pNGB) which is produced massless due to the spontaneous $U(1)_{\rm{PQ}}$ symmetry breaking at a scale $f_a$, called the axion decay constant. It gets mass due to non perturbative Quantum Chromodynamics (QCD) effects at a scale $\Lambda_{\rm{QCD}}$. These are called QCD axions. There are other Axion Like Particles (ALPs) which are not exactly the QCD axions, but have similar properties like the QCD axions. These ALPs are motivated from string compactifications. For ALPs, the axion mass $(m_a)$ and decay constant $(f_a)$ are independent of each other. There are several laboratory, astrophysical, and cosmological experiments which put bounds on $m_a$ and $f_a$.

The SM of particle physics is a gauge theory of $SU(3)_c\times SU(2)_L\times U(1)_Y$ that remains invariant under four global symmetries corresponding to the lepton numbers of the three lepton families and the baryon number. These symmetries are called the accidental symmetries. The conservation of baryon number implies that the proton is stable and the conservation of lepton number demands that the nature of neutrino is Dirac type. One can construct three combinations of these four global symmetries in an anomaly free way and they can be gauged in the SM. These are $U(1)_{B-L}$, $U(1)_{L_e-L_\mu}$, and $U(1)_{L_\mu-L_\tau}$ ($U(1)_{L_e-L_\tau}$ is a linear combination of $U(1)_{L_e-L_\mu}$ and $U(1)_{L_\mu-L_\tau}$). The gauge bosons associated with these symmetries have interesting phenomenology and one can constrain the gauge boson mass and coupling from different experiments. Such $U(1)$ gauge theories can explain several BSM physics such as neutrino mass, dark matter etc. If the mass of the gauge boson is very small, then it can mediate long range force, where the range of the force is determined by the inverse of the gauge boson mass. The $L_i-L_j$
type of long range force can also act as a fifth force and the studies of fifth force can provide complementary checks of such particle physics models.

The ultralight axion and gauge boson can also serve as Fuzzy Dark Matter (FDM) whose mass is $\mathcal{O}(10^{-21}~\rm{eV}-10^{-22}~\rm{eV})$. The de Broglie wavelength of such ultralight DM is of the order of the size of a dwarf galaxy.

The paper is organised as follows. In Section \ref{one} we discuss constraints on ultralight axions from compact binary systems. In Section \ref{two} we discuss constraints on axionic fuzzy dark matter from light bending and Shapiro time delay. The vector gauge boson radiation from compact binary systems in a gauged $L_\mu-L_\tau$ scenario has been discussed in Section \ref{three}. Lastly, in Section \ref{four} we discuss constraints on long range force from perihelion precession of planets in a gauged $L_e-L_{\mu,\tau}$ scenario. 

We have used natural system of units $\hbar=1$ and $c=1$ throughout the paper. 
\section{Constraints on ultralight axions from compact binary systems}\label{one}
If a compact star such as neutron star (NS) or white dwarf (WD) is immersed in a low mass axionic potential and if axions have coupling with nucleons then it develops a long range axion field outside of the star. This axion hair is radiated away when the compact stars are in a binary orbit. The orbital period of compact binary systems (NS-NS, NS-WD etc.) primarily decreases due to the gravitational wave radiation which matches quite well with Einstein's general relativistic prediction. If compact stars contain a substantial amount of axion charge, then the axionic dipole radiation from  the compact binary system can also contribute to the orbital period loss of binary systems. However, the contribution of axion radiation should be within the experimental uncertainty. 
The mass of the radiated axion is constrained by the semi major axis of the binary system, and hence, we obtain $m_a\lesssim 10^{-19}~\rm{eV}$. This mass range is in the ballpark of FDM. 
We consider four compact binary systems (two NS-NS, and two NS-WD), and comparing with the experimental values of orbital period loss, we obtain bounds on $f_a$. In Table \ref{tableeI} we consider four compact binary systems and obtain upper bounds on $f_a$. Here, $\alpha$ denotes the ratio of the axion mediated Yukawa type fifth force to the gravitational force.  
\begin{table}[h]
\caption{\label{tableeI} Upper bounds on $f_a$ for NS-NS (PSR B1913+16, PSR J0737-3039) and NS-WD (PSR J0348+0432, PSR J1738+0333) binary systems. Here we consider the mass of the axions as $m_a\lesssim 10^{-19}~\rm{eV}$.}
\centering
\begin{tabular}{ lcc  }
 
 \hline
Compact binary system & $f_a$ (GeV) & $\alpha$\\
 \hline
PSR J0348+0432  & $\lesssim 1.66\times 10^{11}$  & $\lesssim 5.73\times 10^{-10}$ \\
 PSR J0737-3039 & $\lesssim   9.76\times 10^{16}$  &$\lesssim 9.21\times 10^{-3}$ \\
 PSR J1738+0333 & $\lesssim 2.03\times 10^{11}$  & $\lesssim 8.59\times 10^{-10}$\\
PSR B1913+16 & $\lesssim 2.12\times 10^{17}$  & $\lesssim 3.4\times 10^{-2}$ \\
 \hline
\end{tabular}
\end{table}
We obtain stronger bound on $f_a$ from NS-WD binary system as $f_a\lesssim \mathcal{O}(10^{11}~\rm{GeV})$ \cite{KumarPoddar:2019jxe}. This implies if ALPs are FDM then they do not couple with quarks.  
\section{Constraints on axionic fuzzy dark matter from light bending and Shapiro time delay}\label{two}
\begin{figure}
\centering
\includegraphics[width=3in,angle=360]{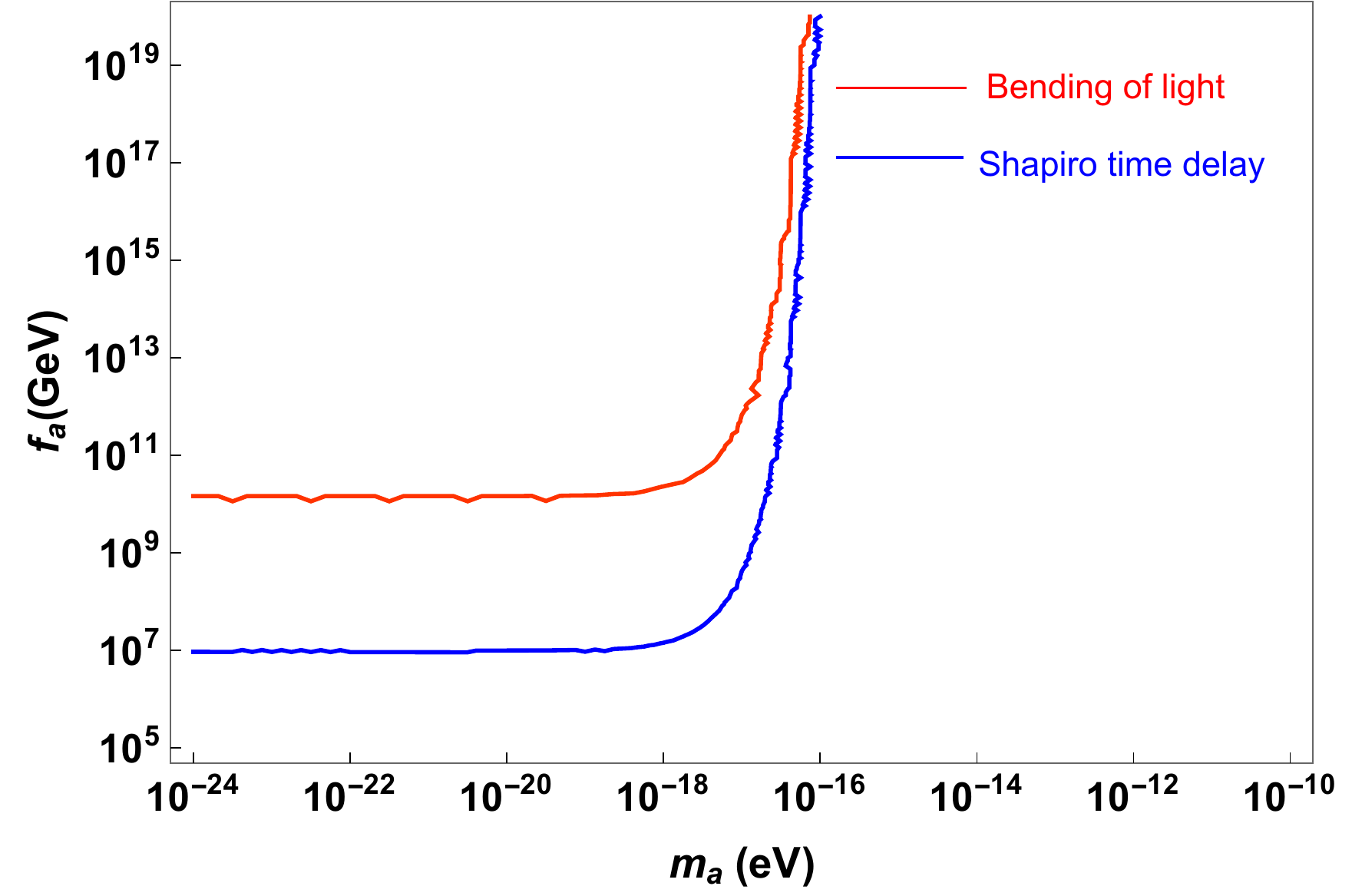}
\caption{Variation of $f_a$ with $m_a$ from light bending and Shapiro delay measurements.}
\label{fig:bending}
\end{figure}

Like NS and WD, if celestial objects like planets and Sun are immersed in a low mass axionic potential and if axions have couplings with nucleons then these objects can also emit long range axion hair. Besides the general relativistic effect, the long range axionic Yukawa potential between Earth and Sun contribute to the measurement of gravitational light bending and Shapiro time delay. However, the contribution of axionic fifth force should be within the experimental uncertainty. 
Comparing with the experimental results, we obtain bounds on axion decay constant as $f_a\lesssim 1.58\times 10^{10}~\rm{GeV}$ from light bending, and $f_a\lesssim 9.85\times 10^{6}~\rm{GeV}$ from Shapiro time delay \cite{Poddar:2021sbc}. The mass of the axion is constrained by the distance between Earth and Sun which gives $m_a\lesssim 10^{-18}~\rm{eV}$. In Figure.\ref{fig:bending} we obtain variation of $f_a$ with the axion mass. 
We obtain stronger bound from Shapiro time delay, and it implies if axions are FDM then they do not couple with quarks.
\section{Vector gauge boson radiation from compact binary systems in a gauged $L_\mu-L_\tau$ scenario}\label{three}
The NS contains lots of muon charge $(\approx 10^{55})$ due to large chemical potential of degenerate electrons, and hence, long range $L_\mu-L_\tau$ type of force can mediate between the stars of compact binary system. The gauge boson of $L_\mu-L_\tau$ type can radiate from the binary system if the mass $(M_{Z^\prime})$ of the gauge boson is smaller than the orbital frequency $(\Omega)$ of the compact binary system. This demands $M_{Z^\prime}\lesssim 10^{-19}~\rm{eV}$. The radiation of ultralight vector gauge bosons can contribute to the orbital period loss of compact binary systems together with the gravitational radiation. However, the contribution of ultralight vector particles should be within the experimental uncertainty limit. We consider four compact binary systems (similar to Section \ref{one}) and comparing with the experimental data we obtain bounds on gauge coupling $(g)$.  
\begin{table}[h]
\caption{\label{tableI} Upper bounds on $g$ for four compact binary systems from fifth force and orbital period decay. We consider $M_{Z^\prime}\lesssim 10^{-19}~\rm{eV}$. }
\centering
\begin{tabular}{ lcc  }
 
 \hline
Compact binary system \hspace{0.5cm} & $g$(fifth force)\hspace{0.5cm} & $g$(orbital period decay)\\
 \hline
PSR B1913+16  & $\leq 4.99\times 10^{-17}$  & $\leq 2.21\times 10^{-18}$ \\
PSR J0737-3039 & $\leq 4.58\times 10^{-17}$  &$ \leq 2.17\times 10^{-19}$\\
PSR J0348+0432 & $ -$  & $\leq 9.02\times 10^{-20}$ \\
PSR J1738+0333 & $ -$  &$ \leq 4.24\times 10^{-20}$\\
 \hline
\end{tabular}
\end{table}
In Table \ref{tableI} we obtain bounds on gauge coupling from orbital period loss of comapct binary systems and fifth force. We do not obtain any bounds on $g$ from NS-WD binary system becuase the WD does not contain any muon charge. 
\begin{figure}[!htbp]
\centering
\includegraphics[width=3.0in,angle=360]{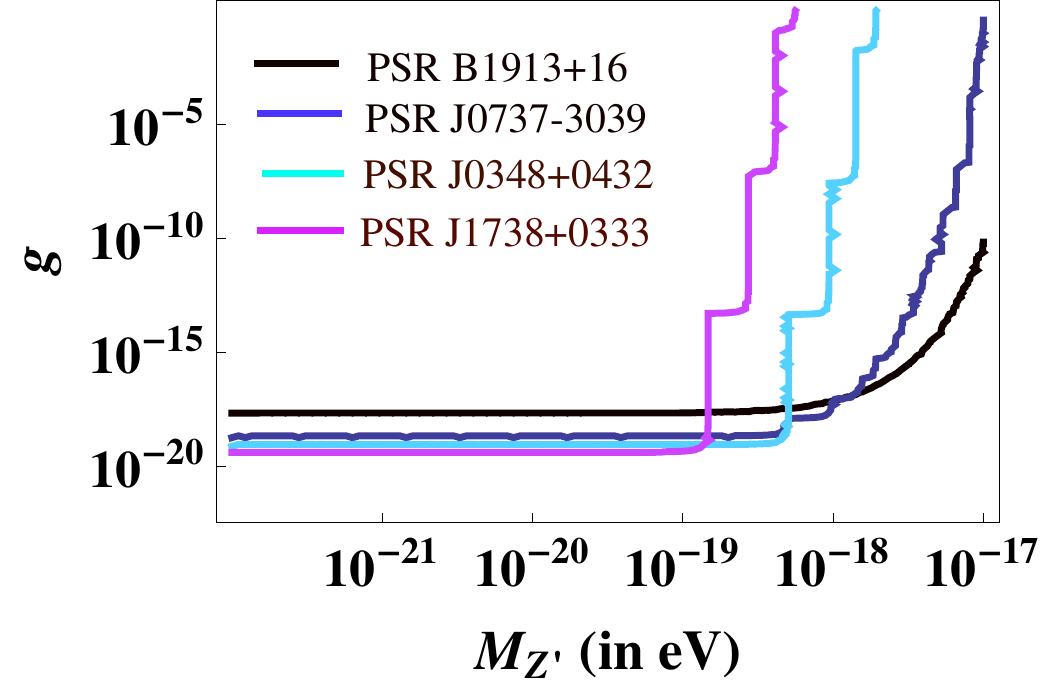}
\caption{Variaton of $g$ vs. $M_{Z^\prime}$ from orbital period loss in a gauged $L_\mu-L_\tau$ scenario.}
\label{fig:PSRJ}
\end{figure} 
In Figure.\ref{fig:PSRJ} we show the variation of gauge coupling with the mass of the gauge boson in a gauged $L_\mu-L_\tau$ scenario. We obtain stronger bound on $g$ as $g\leq 4.24\times 10^{-20}$ from PSR J1738+0333 \cite{KumarPoddar:2019ceq}. These gauge bosons can be candidates for ultralight vector dark matter.
\section{Constraints on long range force from perihelion precession of planets in a gauged $L_e-L_{\mu,\tau}$ scenario}\label{four}
Due to the presence of large number of electrons in Sun and planets, long range Yukawa type $L_e-L_{\mu,\tau}$ force can mediate between the planets and the Sun. The gauge boson of $L_e-L_{\mu,\tau}$ type can mediate between the Sun and planets and contribute to the perihelion precession of planets together with general relativistic prediction. However, the contribution of this Yukawa type fifth force in perihelion precession measurement should be within the experimental uncertainty limit. The mass of the gauge boson is constrained by the distance between Sun and planets which demands $M_{Z^\prime}\lesssim 10^{-19}~\rm{eV}$.
\begin{table*}
\caption{\label{tableperiII} Summary of the upper bounds on $g$ from perihelion precession of planets.}
\label{sphericcase}
\begin{tabular*}{\textwidth}{@{\extracolsep{\fill}}lcc@{}}
\hline
Planet & \multicolumn{1}{c}{Uncertainty in perihelion advance (as/cy)} & \multicolumn{1}{c}{$g$ from perihelion advance}\\
\hline
Mercury  & $3.0\times 10^{-3}$  & $\leq1.055\times 10^{-24}$\\
Venus & $1.6\times 10^{-3}$  &$\leq1.377\times 10^{-24}$\\
Earth & $1.9\times 10^{-4}$  & $\leq6.021\times 10^{-25}$ \\
Mars &  $3.7\times 10^{-5}$  &$\leq3.506\times 10^{-25}$\\
Jupiter & $2.8\times 10^{-2} $  &$\leq2.477\times 10^{-23}$\\
Saturn & $4.7\times 10^{-4}$  &$\leq5.040\times 10^{-24}$\\
\hline
\end{tabular*}
\end{table*}
In Table \ref{tableperiII} we obtain bounds on gauge coupling $(g)$ from perihelion precession of planets. 
\begin{figure*}[!htbp]
\centering
\includegraphics[width=3.0in,angle=360]{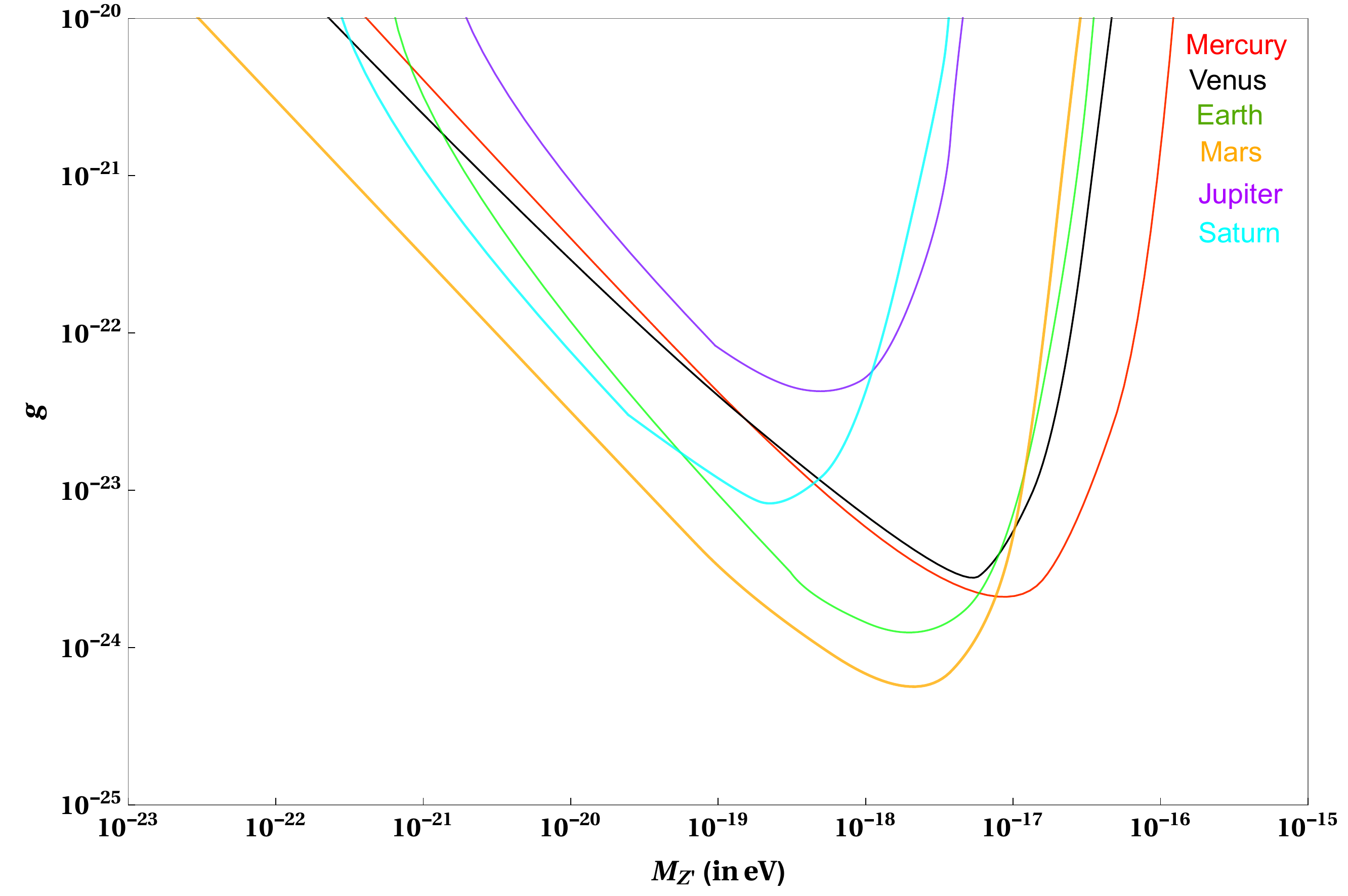}
\caption{Variation of gauge coupling $(g)$ with the mass of gauge boson $M_{Z^\prime}$.}
\label{fig:planet}
\end{figure*}
In Figure.\ref{fig:planet} we obtain bounds on gauge coupling for planets upto Saturn. The planet Mars puts the stronger bound on $g$ \cite{KumarPoddar:2020kdz}. These gauge bosons can also be candidates for ultralight vector dark matter.
\section{Conclusion}
In this paper we consider ultralight particles (axions and vector gauge bosons) and their possible searches from astrophysical observations and experiments. We obtain bounds on axion decay constant from orbital period loss of compact binary systems (indirect evidence of gravitational waves), gravitational light bending, and Shapiro time delay. The Shapiro time delay puts stronger bound on axion decay constant $(f_a\lesssim \mathcal{O}(10^7)~\rm{GeV})$. It concludes if ultralight ALPs are FDM, then they do not couple with quarks. We also obtain upper bounds on gauge coupling of $L_i-L_j$ scenario from orbital period loss of compact binary systems and perihelion precession of planets. We obtain upper bound on the gauge coupling as $g\lesssim \mathcal{O}( 10^{-20})$ for $L_\mu-L_\tau$ scenario from orbital period loss of compact binary systems. We also obtain the upper bound on gauge coupling as $g\lesssim \mathcal{O}(10^{-25})$ for $L_e-L_{\mu,\tau}$ scenario from perihelion precession of planets. Such ultralight vector particles can be promising candidates for DM. These bounds are the complementary checks for such particle physics models.

\bibliographystyle{utphys}
\bibliography{ichep}
\end{document}